\documentclass[twocolumn,superscriptaddress,amsmath,amssymb,showpacs,prb]{revtex4-2}

\usepackage{graphicx}
\usepackage{color}
\usepackage[colorlinks=true,citecolor=blue, urlcolor = blue, linkcolor= blue]{hyperref}
\renewcommand{\phi}{\varphi}

\begin{document}

\title{Electron-phonon coupling strength from \textit{ab initio} frozen-phonon approach}

\author{Yang Sun}
	\affiliation{Department of Physics and Ames Laboratory, Iowa State University, Ames, IA 50011, USA}
	\affiliation{Department of Applied Physics and Applied Mathematics, Columbia University, New York, NY 10027, USA}

\author{Feng Zhang}
    \affiliation{Department of Physics and Ames Laboratory, Iowa State University, Ames, IA 50011, USA}

\author{Cai-Zhuang Wang}
    \affiliation{Department of Physics and Ames Laboratory, Iowa State University, Ames, IA 50011, USA}

\author{Kai-Ming Ho}
    \affiliation{Department of Physics and Ames Laboratory, Iowa State University, Ames, IA 50011, USA}
\author{Igor I. Mazin}
    \affiliation{Department of Physics and Astronomy and Quantum Science and Engineering Center, George Mason University, Fairfax, Virginia 22030, USA}

\author{Vladimir Antropov}
	\affiliation{Department of Physics and Ames Laboratory, Iowa State University, Ames, IA 50011, USA}

\begin{abstract}

We propose a fast method for high-throughput screening of potential superconducting materials. The method is based on calculating metallic screening of zone-center phonon modes, which provides an accurate estimate for the electron-phonon coupling strength. This method is complementary to the recently proposed Rigid Muffin Tin (RMT) method, which amounts to integrating the electron-phonon coupling over the entire Brillouin zone (as opposed to the zone center), but in a relatively inferior approximation. We illustrate the use of this method by applying it to MgB$_\text{2}$, where the high-temperature superconductivity is known to be driven largely by the zone-center modes, and compare it to a sister compound AlB$_\text{2}$. We further illustrate the usage of this descriptor by screening a large number of binary hydrides, for which accurate first-principle calculations of electron-phonon coupling have been recently published. Together with the RMT descriptor, this method opens a way to perform initial high-throughput screening in search of conventional superconductors via machine learning or data mining.

\end{abstract}

\maketitle

One of the most rapidly developing areas of computational materials science is the theoretical search for functional materials using such information technologies as data mining, machine learning and artificial intelligence  \cite{1,2,3,4,5}. At the crux of this field is the ability to screen candidates not only for stability (which is more or less routine now) but, most importantly, for a particular figure of merit. In some cases, testing material for the property of interest is relatively inexpensive, and such cases have been intensively investigated in the last decade. Typical examples include topological bands  \cite{6,7}, thermoelectric properties  \cite{8}, and battery materials  \cite{9}. On the other hand, useful properties that require tedious, often poorly converged and time-consuming calculations are naturally not in the first row of this event. Still, as the low-hanging fruits are being exhausted, researchers’ attention is turning toward those in the latter category.

One of the relatively less explored, due to this reason, parts of the landscape, is the massive computational search for conventional superconductors. While the computational algorithm based on the linear response formalism in the DFT is well known, it is very time-consuming, even with reduced accuracy. When scanning hundreds and thousands of candidates, it is helpful to have a fast, albeit less accurate method, to “enrich” the sampling space the same way as miners enrich ores: to weed out the candidates that are \textit{unlikely} to have strong electron-phonon coupling (albeit this is not guaranteed), and elevate to the next stage those that are likely to have one (albeit many will not).

To this end, a half a century-old Rigid Muffin Tin (RMT) method has been revived  \cite{10}. This method calculates the integrated electron-phonon coupling (EPC) constant exactly, under one very important assumption that the crystal space can be partitioned into non-overlapping spheres (MT spheres) and interstitial space, and the crystal potential in the spheres is spherically symmetric and that in the interstitial regions is constant. Furthermore, it is assumed that as ions are displaced from their equilibrium positions, the potential inside each MT sphere shifts rigidly. This method is extremely fast and gives reasonably accurate results for close-packed transition metals, where the potential is mainly due to well-localized \textit{d}-electrons (e.g., Nb, V, Mo)  \cite{11}, but strongly underestimates the coupling constant for \textit{sp}-metals (Al, Pb, MgB$_\text{2}$)  \cite{12}. Below we offer another method that works very well in MgB$_\text{2}$ and reasonably well in hydride systems, which is based on calculating the EPC for selected phonons (specifically, zone-center modes), and then making an \textit{ad hoc} assumption that these modes are reasonably representative for the total EPC. Since the underlying assumptions are totally different from those in the RMT method, we suggest they can be used complementarily. 

\begin{figure}
\includegraphics[width=0.48\textwidth]{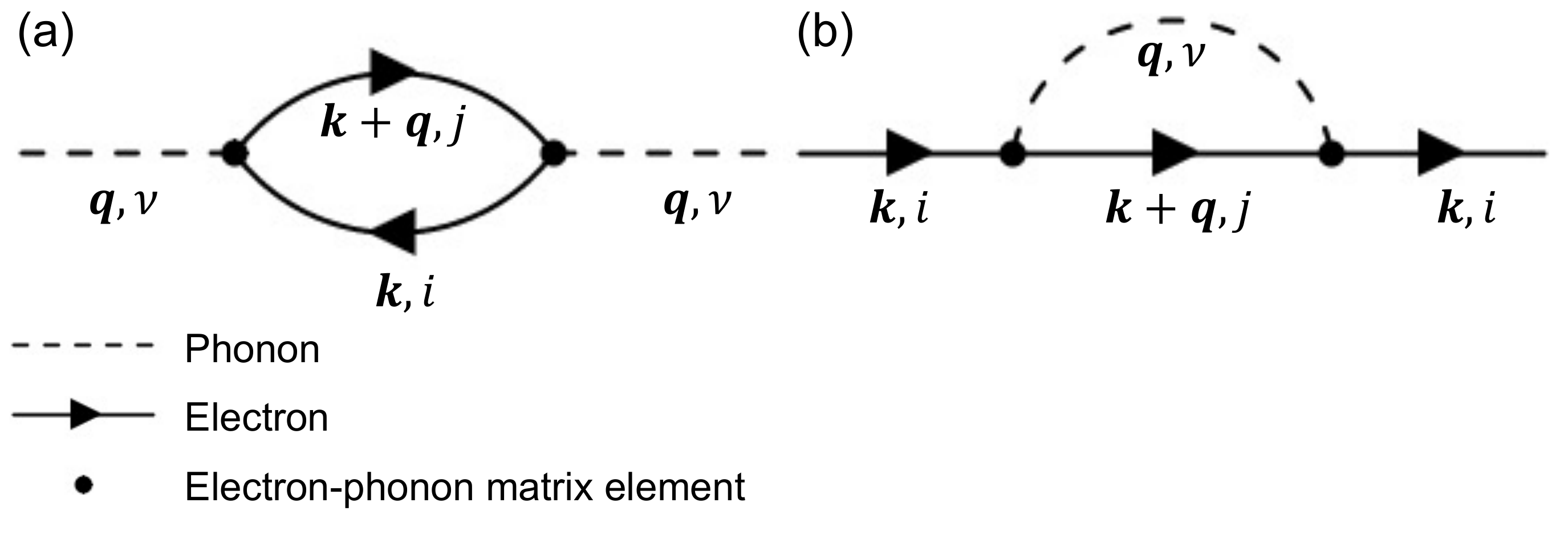}
\caption{\label{fig:fig1} Feynmann diagrams for electron and phonon self energies.}
\end{figure}

The fact that phonon softening is controlled by the same physics as superconducting EPC comes from the simple observation that the former is the real part of the phonon self-energy $\delta \omega^{2}$, while the latter is determined by the imaginary part of the same quantity, $\lambda=\sum_{\boldsymbol{q} v} \frac{\gamma_{\boldsymbol{q} v}}{\pi N\left(\varepsilon_{F}\right) \omega_{\boldsymbol{q} v}^{2}}$ is the phonon linewidth, $N(\varepsilon_F)$ is the density of states at the Fermi level. This can be visually appreciated from the Feynman diagrams for the phonon (Fig.~\ref{fig:fig1}a) and electron (Fig.~\ref{fig:fig1}b) self-energy. The former gives the softening (real part) and the broadening (imaginary part) of the phonon line, the latter is the second Eliashberg diagram, so its real part gives electron mass renormalization and the imaginary part, after summation over all the phonon modes, the anisotropic Eliashberg function  \cite{13}

In Ref.  \cite{14} a simple expression was derived for both quantities, in the special case of $q \rightarrow 0$, in terms of the electron-phonon matrix element $g$ (the dot in Fig. 1). In the following we recap this picture. The one-electron contribution to the dynamical matrix is defined as $D=\frac{d^2E_1}{dR^2}$ where $R$ is the ionic coordinate. For simplicity we consider a phonon with just a single atom displaced. Then we can define $g$ as $g=\frac{1}{\sqrt{2M\omega}}\frac{\partial\varepsilon_{i \boldsymbol{k}}} {\partial R}$, where $\omega$ is the full self-consistent frequency, i.e., the fully screened phonon frequency. The total energy (the chemical potential is set to zero) is $E_1=\sum_{\sigma i\boldsymbol{k}}\varepsilon_{\sigma i\boldsymbol{k}}\theta(-\varepsilon_{\sigma i\boldsymbol{k}})=2\sum_{i\boldsymbol{k}}\varepsilon_{i\boldsymbol{k}}\theta(-\varepsilon_{i\boldsymbol{k}})$, where $\varepsilon_{i\boldsymbol{k}}$ is the quasiparticle energy and $\theta(-\varepsilon_{\sigma i \boldsymbol{k} })$ is the Heaviside’s theta function, and the Fermi energy is set to zero. This leads to

\begin{equation}
\frac{d^{2} E_{1}}{d R^{2}}=2 \sum_{i \boldsymbol{k}}\left[\frac{d^{2} \varepsilon_{i \boldsymbol{k}}}{d R^{2}} \theta\left(-\varepsilon_{i \boldsymbol{k}}\right)-2 \delta\left(\varepsilon_{i \boldsymbol{k}}\right)\left(\frac{d \varepsilon_{i \boldsymbol{k}}}{d R}\right)^{2}\right]
\label{eq1}.
\end{equation}

Here summations over $\boldsymbol{k}$ are normalized to one state per Brillouin zone (BZ). The first term here can be interpreted as an unscreened dynamic matrix where the potential is not affected by the electron-phonon interactions. We define it as $\widetilde{D}=2\sum_{i\boldsymbol{k}}{\frac{d^2\varepsilon_{i\boldsymbol{k}}}{dR^2}\theta\left(-\varepsilon_{i\boldsymbol{k}}\right)}$ and its corresponding frequency as an unscreened phonon frequency, i.e. $\widetilde{D}=M{\widetilde{\omega}}^2$. Replacing the second term with $g$, we get

\begin{equation}
M\omega^2=M{\widetilde{\omega}}^2-4\sum_{i\boldsymbol{k}}{2M\omega\left|g\right|^2}\delta\left(\varepsilon_{i\boldsymbol{k}}\right)
\label{eq2}.
\end{equation}

Therefore,

\begin{equation}
\frac{\omega^2-{\widetilde{\omega}}^2}{\omega^2}=\frac{8}{\omega}\sum_{i\boldsymbol{k}}\left|g\right|^2\delta\left(\varepsilon_{i\boldsymbol{k}}\right)
\label{eq3}.
\end{equation}

The definition of EPC constant per mode   \cite{13} in the Eliashberg theory is $\lambda_{\boldsymbol{q}}=\frac{2}{N(\varepsilon_F)\omega_{\boldsymbol{q}} } \sum_{ij\boldsymbol{k}} |g| ^2 \delta (\varepsilon_{i\boldsymbol{k}}) \delta (\varepsilon_{i\boldsymbol{k}+\boldsymbol{q}} - \varepsilon_{j\boldsymbol{k}} - \omega_{\boldsymbol{q}} )$ and, after integrating over the entire Brillouin zone, $\lambda_{\text{BZ}}=\sum_{\boldsymbol{q}}\lambda_{\boldsymbol{q}}$. Note that this expression is zero at $q=0$, reflecting the Landau threshold. However, assuming an Einstein mode, i.e., $g$ independent on $\boldsymbol{q}$, setting $\omega$ to zero for finite $\boldsymbol{q}$, as it is customary  \cite{13}, and integrating over $\boldsymbol{q}$, we get

\begin{equation}
\lambda_{\text{BZ}}=\frac{2}{\omega}\sum_{ij\boldsymbol{k}}\left|g\right|^2\delta\left(\varepsilon_{i\boldsymbol{k}}\right)
\label{eq4}.
\end{equation}

Equation~(\ref{eq3}) is similar to the one in Eliashberg theory (Eq.~\ref{eq4}), except for different prefactors. If we define $\lambda_\Gamma=\frac{\omega^2-{\widetilde{\omega}}^2}{4\omega^2}$, then we can assume that ${\lambda_{\text{BZ}}=f\lambda}_\Gamma$, where $f=1$ within the assumptions made. In this formalism, $f$ accounts for the phase-space factor, i.e., the difference between the zone-center calculations and the proper integration over all phonons.

Our main hypothesis is that this “fudge factor” is reasonably constant within the same family of materials. In the following, we will use a few cases to examine this hypothesis. The two frequencies $\omega$ and $\widetilde{\omega}$ can be calculated in any DFT package using the frozen phonon method, with the difference that in one case we calculate the fully self-consistent total energies, while in the other the occupation numbers for the electronic states are kept fixed with the frozen phonon displacement. 

Density functional theory (DFT) calculations in the following were carried out using the projector augmented wave (PAW) method  \cite{15} implemented in the VASP code  \cite{16,17}. The exchange and correlation energy are treated with the generalized gradient approximation (GGA) and parameterized by the Perdew-Burke-Ernzerhof formula (PBE)  \cite{18}. A plane-wave basis was used with a kinetic energy cutoff of 520 eV, and the convergence criterion for the total energy was set to $10^{-8}$ eV. The $\Gamma$-centered Monkhorst-Pack grid was adopted for Brillouin zone sampling. The $\boldsymbol{k}$-point mesh was generated based on the length parameter $R_{\boldsymbol{k}}$ as $N_{\boldsymbol{k}}=R_{\boldsymbol{k}}\left|\boldsymbol{b}\right|$, where $\boldsymbol{b}$ is the reciprocal lattice vectors. Different $\boldsymbol{k}$-point meshes were tested in the phonon calculations. The zone-center phonons were calculated by the frozen-phonon method with finite differences. The displacement amplitude in the frozen-phonon calculations is 0.02 $\text{\AA}$. 

The screened phonon frequency $\omega$ was computed by fully self-consistent (SCF) calculations in the displaced atomic configurations using the tetrahedron method with Bl$\text{\"{o}}$chl corrections (ISMEAR = -5). To compute the unscreened phonon frequency $\widetilde{\omega}$, an SCF calculation with the tetrahedron method was first performed in the equilibrium configuration, followed by the SCF calculations with the displaced atoms, but with partial occupations fixed at the equilibrium configuration (ISMEAR = -2, i.e., the occupation number are read from the stored WAVECAR file and never recalculated). A caveat here is that the point symmetry and the size of the inequivalent wedge of the BZ change upon imposing a frozen phonon (unless an $\text{A}_{\text{1g}}$ phonon). The simplest but the least efficient way to handle it is to turn off the symmetry entirely and use the built-in frozen-phonon capability in VASP (IBRION=5 or 6, ISYM=0). However, in that case, VASP does not perform the symmetry analysis of the phonon modes either, so instead of calculating only ionic displacements compatible with the given phonon representation, it generates all $3N(3N+1)/2$ displacement patterns, where $N$ is the number of atoms in the unit cell. Alternatively, one can use an external software suite, such as PHONOPY \cite{19} or SMODES (part of the ISOTROPY suite)  \cite{20} to generate only the relevant displacement patterns and perform frozen-phonon calculations separately for each representation.  A further speed-up can be achieved by calculating only the Raman-active modes, because otherwise, they would not have any EPC by symmetry. The SMODES program provides this information, too. Depending on the size of the unit cell and the symmetry of the crystal, the saving in computer time may be up to an order of magnitude or even more. Finally, one can save even more time by calculating the occupation numbers in the highest (undistorted) symmetry and using the point group operations to repeat them into the rest of the BZ. However, this would require some modification of the VASP source code. In the calculations presented below we used an intermediate protocol, utilizing PHONOPY to generate displacements for each irreducible representation, and VASP to compute the screened and unscreened phonon frequencies.

\onecolumngrid

\begin{figure}[b]
\includegraphics[width=0.7\textwidth]{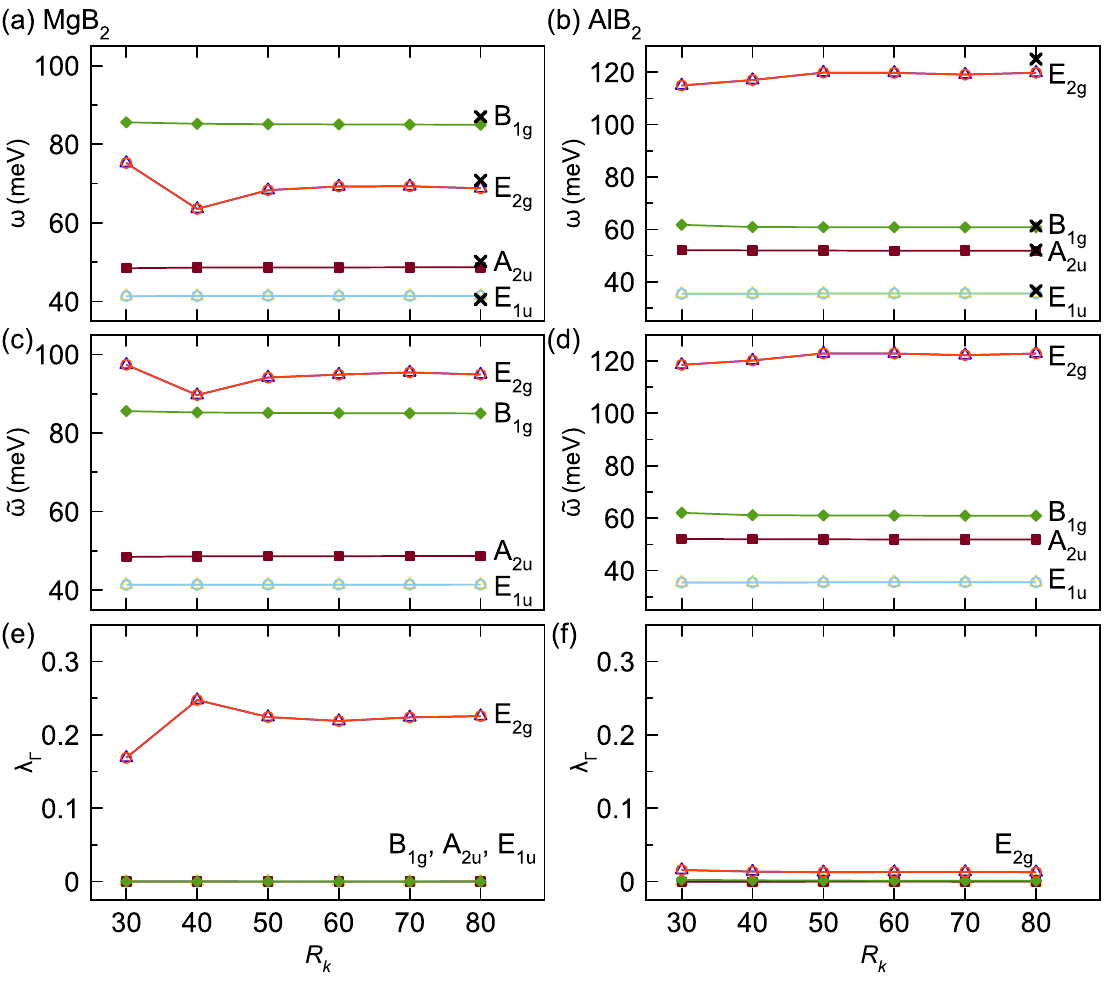}
\caption{\label{fig:fig2} Zone-center EPC strength calculations for MgB$_\text{2}$ and AlB$_\text{2}$. (a)-(b) the screened phonon frequencies $\omega$. (c)-(d) unscreened phonon frequencies $\widetilde{\omega}$. (e)-(f) zone-center EPC strength $\lambda_\Gamma$. $R_{\boldsymbol{k}}$ is the reciprocal lattice spacing that determines the k grid density. The reference data marked by ‘$\times$’ in (a) and (b) are from \cite{29}.}
\end{figure}

\twocolumngrid

\textit{MgB$_\text{2}$ and AlB$_\text{2}$}.\textemdash We first test the method by computing the zone-center EPC strength of the MgB$_\text{2}$ and AlB$_\text{2}$. Both phases adopt the same “AlB$_\text{2}$-type” crystal structure (P6/mmm). MgB$_\text{2}$ has a high $T_c$ of 39 K \cite{21} and strong EPC interactions \cite{22,23,24,25,26,27,28}, while AlB$_\text{2}$ shows no superconductivity in experiments and weak EPC \cite{29}. Figures~\ref{fig:fig2}(a) and (b) show the frequencies of optical phonon modes at the zone center computed by the frozen-phonon method for MgB$_\text{2}$ and AlB$_\text{2}$. The calculations are tested with different $\boldsymbol{k}$-points grids and found to converge at $R_{\boldsymbol{k}}=60$ (i.e. $22\times22\times17$). The computed phonon frequencies of both MgB$_\text{2}$ and AlB$_\text{2}$ agree well with previous calculations in \cite{29}. The unscreened phonon frequency $\widetilde{\omega}$ are computed for MgB$_\text{2}$ and AlB$_\text{2}$, shown in Fig.~\ref{fig:fig2}(c) and (d), respectively. Comparing the screened and unscreened phonon frequencies of MgB$_\text{2}$ in Fig.~\ref{fig:fig2}(a) and (c), one can see that the $\text{E}_{\text{2g}}$ modes, which are the in-plane boron stretching modes, show a strong softening due to the screening of electron-phonon interactions, while other modes remain unchanged. This is consistent with previous studies that the softening of $\text{E}_{\text{2g}}$ mode is the main contribution to the strong EPC of MgB$_\text{2}$ \cite{30,31}. In AlB$_\text{2}$, there is almost no difference between screened and unscreened phonon frequencies. Figures~\ref{fig:fig2}(e) and (f) show the zone-center EPC strength $\lambda_\Gamma$ for each phonon mode. In MgB$_\text{2}$, the $\text{E}_{\text{2g}}$ mode shows a $\lambda_\Gamma$ of 0.23, while other optical modes, as expected by symmetry, show no EPC. In AlB$_\text{2}$, the $\text{E}_{\text{2g}}$ mode also shows a non-zero EPC strength while the amplitude is one order of magnitude smaller than the one in MgB$_\text{2}$. Other modes in AlB$_\text{2}$ also do not contribute to the EPC. These results demonstrate that the simple frozen-phonon calculations of screened and unscreened phonon modes provide a correct physical picture of the electron-phonon interactions and a qualitative estimate of EPC strength in MgB$_\text{2}$ and AlB$_\text{2}$. Therefore, this test provides a good validation of the current method. We note that the computational cost of these frozen-phonon calculations is very small, e.g., it only takes $\sim$20 minutes to compute $\lambda_\Gamma$ for MgB$_\text{2}$ with $R_{\boldsymbol{k}}=60$ on a 32-core Intel(R) Xeon(R) Gold 6130 CPU.

\begin{figure}[b]
\includegraphics[width=0.78\textwidth]{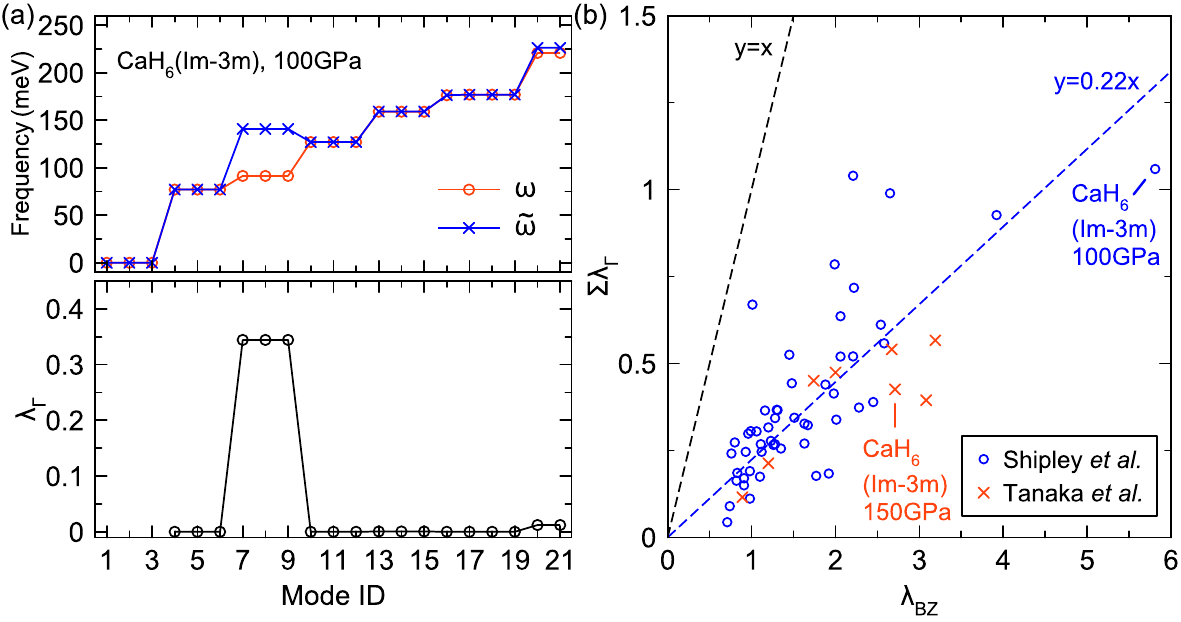}
\caption{\label{fig:fig3} (a) The screened and unscreened phonon frequencies (top panel) and zone-center EPC strength (bottom panel) in the CaH$_\text{6}$ (Im$\bar{\text{3}}$m) phase at 100 GPa. (b) Scatter plot of zone-center EPC strength $\lambda_\Gamma$ versus EPC constant from full Brillouin zone calculations $\lambda_{\text{BZ}}$ for datasets by Tanaka \textit{et al.} \cite{32} and Shipley \textit{et al.} \cite{33}. The dashed lines indicate linear relations with different slopes. }
\end{figure}

\textit{High-pressure hydrates}.\textemdash We apply the method to compute the zone-center EPC strength for two datasets of high-pressure hydrates developed in Ref. \cite{32,33}. The dataset from Ref.  \cite{32} contains eight systems intensely studied in the last decades. The dataset from Ref.  \cite{33} predicts 52 hydrate systems in the pressure range from 100-500GPa, with some confirmed experimentally. The full Brillouin zone EPC constant, $\lambda_{\text{BZ}}$, has been calculated in both datasets \cite{32,33}, which provides a perfect reference to investigate the relation between zone-center EPC strength $\lambda_\Gamma$ and EPC constant $\lambda_{\text{BZ}}$. We computed $\lambda_\Gamma$ for all these phases and take the CaH$_\text{6}$ (Im$\bar{\text{3}}$m) phase as an example. CaH$_\text{6}$ (Im$\bar{\text{3}}$m) phase was first predicted to have strong EPC and large $T_c$ at high pressures  \cite{34} and was recently synthesized in experiments \cite{35}. In Fig.~\ref{fig:fig3}(a), we computed the screened and unscreened zone-center phonon frequencies for CaH$_\text{6}$ (Im$\bar{\text{3}}$m) phase at 100 GPa. The triple degenerate modes at $\sim$100 meV show a strong softening due to EPC. The double degenerate modes at $\sim$220 meV also show a slight softening. Therefore, these modes provide the main contribution to the zone-center EPC, which can be seen in Fig.~\ref{fig:fig3}(b). This is qualitatively consistent with the full Brillouin zone EPC  \cite{34}. The triple modes yield  $\lambda_\Gamma=0.34$. The summation of zone-center EPC of all modes noted as $\sum\lambda_\Gamma$, is 1.06. According to the calculations in \cite{33}, the full Brillouin zone EPC constant $\lambda_{\text{BZ}}$ of CaH$_\text{6}$ (Im$\bar{\text{3}}$m) phase at 100 GPa was 5.81. The difference between $\sum\lambda_\Gamma$ and $\lambda_{\text{BZ}}$ is attributed mainly to the zone-boundary phonons in CaH$_\text{6}$ (Im$\bar{\text{3}}$m), e.g., at H, N and P points, as shown in Ref. \cite{34}. The EPC of CaH$_\text{6}$ has a strong pressure dependence \cite{33}. When the pressure increases to 150 GPa, $\lambda_{\text{BZ}}$ significantly drops to 2.71  \cite{32}. This is also captured by $\sum\lambda_\Gamma$, which decreases to 0.42 at 150 GPa, shown in Fig.~\ref{fig:fig3}(b).

To compare the zone-center EPC and full Brillouin zone EPC with better statistics, we compute $\sum\lambda_\Gamma$ and plot it with $\lambda_{\text{BZ}}$ for all the systems from Refs. \cite{32,33} in Fig.~\ref{fig:fig3}(b). It shows that $\sum\lambda_\Gamma$ has a positive relation with $\lambda_{\text{BZ}}$. The data can be described by linear regression with a relation of $\sum\lambda_\Gamma\approx0.22\lambda_{\text{BZ}}$, giving the fudge-factor for this system $f\approx0.2$. Some of the data points are even close to the $y=x$ curve, which indicates the zone-center EPC dominants the EPC in these materials. Therefore, the simple zone-center EPC via the frozen phonon calculations can be used as a quick screening for the high EPC without full Brillion zone calculations. 

\onecolumngrid

\newpage

\twocolumngrid

In summary, we suggest using single-cell, $q=0$, frozen-phonon calculations of the EPC strength of the zone-center Raman-active phonons as a quick-n-dirty proxy for full DFT evaluation of the Eliashberg function. The protocol includes one adjustable parameter, which is roughly a constant within one materials family, such as high-pressure hydrides. Test on the AlB$_\text{2}$ and MgB$_\text{2}$ shows that the method can clearly distinguish a promising superconducting material (MgB$_\text{2}$) from a dud (AlB$_\text{2}$). The calculations for the binary hydrate dataset show that this method can be used as a reasonable descriptor of the full Brillouin zone EPC constant across a broad range of materials (from $\lambda\approx0.7$ to nearly 6) within the same family. Therefore, this method can enable a quick large-scale screening for potential high-temperature conventional superconductors. It can also be used in conjunction with other complementary descriptors, such as the RMT method.

\begin{acknowledgments}
\textit{Acknowledgements} We thank Alexey Kolmogorov and Roxana Margine for helpful discussions. This work was supported primarily by National Science Foundation Awards No. DMR-2132666.  C.Z.W. was supported by the US Department of Energy (DOE), Office of Science, Basic Energy Sciences, Materials Science and Engineering Division. Ames Laboratory is operated for the US DOE by Iowa State University under Contract No. DE-AC02-07CH11358. I.I.M. acknowledges support from the National Science Foundation Award No. DMR-2132589.
\end{acknowledgments}

\bibliographystyle{apsrev4-1}

\end{document}